\documentclass[twocolumn,showpacs,amsmath,amssymb,prd,nofootinbib]{revtex4-2}
\usepackage{graphicx}
\usepackage{epsfig}
\usepackage{enumerate}
\def\beq{\begin{equation}}
\def\eeqno#1{\label{#1}\end{equation}}

\def\rarrow{\rightarrow }

\def\dleft{\rlap{{\it D}}\raise 8pt
\hbox{$\scriptscriptstyle\Leftarrow$}}
\def\dright{\rlap{{\it
D}}\raise 8pt\hbox{$\scriptscriptstyle\Rightarrow$}}

\def\kms{{\rm km~s^{-1}}}
\def\cmss{{\rm cm~s^{-2}}}

\def\kpc{~{\rm kpc}}

\def\msun{M_{\odot}}
\def\az{a_{0}}

\def\l0{\ell_{0}}

\def\l{\lambda}

\def\n{\nu}
\def\z{\zeta}

\def\vinf{V\_{\infty}}

\def\av#1{\langle#1\rangle}

\def\xlimin{{x\rarrow\infty \atop{\raise 1pt\hbox to 30pt
{\rightarrowfill}}}}
\def\limlim#1#2{{#1\rarrow #2 \atop{\raise 1pt\hbox to 30pt
{\rightarrowfill}}}}

\def\S{\Sigma}

\def\M{\mathcal{M}}

\def\n{\nu}

\def\_#1{_{\scriptscriptstyle #1}}
\def\^#1{^{\scriptscriptstyle #1}}

\def\mb{\M}
\def\rM{r\_M}
\def\jM{j\_M}

\def\avv{\av{V}_j}
\def\avr{\av{r}}
\def\avS{\av{\S}}

\def\SM{\S\_M}
\begin{document}
\title{MOND fiducial specific angular momentum of disc galaxies}

\author{Mordehai Milgrom}
\affiliation{Department of Particle Physics and Astrophysics, Weizmann Institute, Rehovot 7610001 Israel}

\begin{abstract}
It is pointed out that MOND defines a fiducial specific angular momentum (SAM) for a galaxy of total (baryonic) mass $\mb$: $\jM(\mb)\equiv\mb^{3/4}(G^3/\az)^{1/4}\approx 383(\mb/10^{10}\msun)^{3/4}\kpc~\kms$. It plays important roles in disc-galaxy dynamics and evolution: It underlies scaling relations in virialized galaxies that involve their angular-momentum. I show that the disc SAM should be $j\_D\approx[\avr/\rM(\mb)]\jM(\mb)=[\SM/\avS]^{1/2}\jM(\mb)$, with $\avr$ the mean radius of the disc, $\avS=\mb/2\pi\avr^2$ some mean surface density of the galaxy, $\rM=(\mb G/\az)^{1/2}$ is the MOND radius of the galaxy, and $\SM=\az/2\pi G$ is the (universal) MOND surface density.
So, e.g., for a fixed $\avS$, $j\_D\propto \mb^{3/4}$, while for a fixed $\avr$, $j\_D\propto \mb^{1/4}$.
Furthermore, $\jM(\mb)$ is a reference predictor of the type of galaxy a protogalaxy will settle into, if it evolves in isolation: A protogalaxy of mass $\mb$, and SAM $j\gg\jM(\mb)$ should settle into a low-surface-density disc -- with mean acceleration $\av{a}/\az\approx \jM/j\ll1$. While a protogalaxy with $j\lesssim\jM(\mb)$ should end up with a disc of mass $\mb\_D\approx j\mb/\jM(\mb)$, having a SAM $j\_D\approx\jM(\mb)$, which is tantamount to $\av{a}\approx\az$ (i.e., at the ``Freeman limit''); it should also develop a low-SAM component (e.g., a bulge or a halo), taking up the rest of the mass $\mb\_B\approx\mb-\mb\_D$.
\end{abstract}
\maketitle

\section{Introduction}
Dimensionful constants appearing in a theory enable one to define dimensionful attributes of a system that are of special significance in the dynamics of that system. In quantum mechanics, Planck's $\hbar$ defines the de Broglie wavelength of a particle, given its momentum, and in relativistic quantum mechanics, $\hbar$ (with $c$) defines the Compton wavelength of a particle given its mass. In relativistic gravity, $G$ and $c$ define the Schwarzschild radius of a body, given its mass. All these quantities are of obvious and proven usefulness.
\par
MOND \cite{milgrom83} -- which strives to account for the mass discrepancies in The Universe without dark matter -- introduces a new constant, $\az$, with the dimensions of acceleration (for reviews of MOND see Refs. \cite{fm12,milgrom14,milgrom20}). Two system-specific attributes of proven usefulness that MOND defines with the use of $\az$ (and $G$) are the ``MOND radius'' and the ``MOND speed'' of a given mass.
\par
$\az$, together with $c$ and $G$, also define universal, not-system-specific, quantities, the MOND length, $\ell\_M\equiv c^2/\az$, the MOND time $t\_M\equiv c/\az$, and the MOND mass, $M\_M\equiv c^4/G\az$. Their values are of cosmological magnitudes, which perhaps points to the fundamental basis of MOND (e.g., Ref, \cite{milgrom20a}).
They are analogous to the universal Planck length, time, and mass, which $\hbar$ defines, together with $c$ and $G$.
\par
Here, I point out that MOND also defines an important fiducial value of the specific angular momentum (hereafter, SAM), and I elaborate on the ways in which it enters the dynamics of galaxies and their evolution.
\par
Correlations are observed between global properties of the baryonic body of disc galaxies -- such as their total mass, disc mass, rotational speed, etc. -- and their angular momentum (hereafter, AM) (see, e.g., the recent Refs. \cite{og14,jb19,posti18,fr18,pina21a} for but a few examples). These have given rise to much discussion as to their significance and possible reasons  for their emergence in the dark-matter paradigm (e.g., Ref. \cite{dalcanton97} and Refs. \cite{og14,posti18,fr18,jb19,pina21a}). Reference \cite{lelli19} discusses
AM-mass relations vis a vis the baryonic Tully-Fisher relation.
\par
Correlations involving the AM require, in the dark-matter paradigm, separate explanations, because they hinge on mutual effects of the baryonic and the putative dark-matter halo; e.g., the exchange of AM between the components.
The correlations I specifically discuss here can perhaps be worked by hand into the results of dark-matter simulations, as is usually done in the context of this paradigm. In MOND, however, such correlations follow naturally and without recourse to complicated galaxy formation and evolution scenarios, as MOND is much more constrained than the dark-matter paradigm.
\par
Inasmuch as MOND predicts correctly, as it does, the rotation curves of individual disc galaxies from their baryonic mass distribution, it, perforce, also predicts correctly their AM from the mass distribution. However, this does not directly imply correlations between global properties and the AM. It is thus interesting to inquire how such correlations arise in MOND.

\par
In Sec. \ref{SAM}, I define the MOND fiducial SAM. In Sec. \ref{scaling}, I show how it underlies galaxy correlations involving their AM, and, in Sec. \ref{evolution}, how it is an important predictor of the type of galaxy a protogalaxy will settle to, explaining why some galaxies end up as low-surface-density, pure discs, and some as high-surface-density, disc-plus-bulge ones. In Sec. \ref{observation}, I compare with some relevant observations. Section \ref{discussion} is a discussion of further points.
\section{The MOND specific angular momentum of a galaxy \label{SAM}}
For a body of (``baryonic'') mass $\mb$ (a ``galaxy'' hereafter), the MOND constant, $\az$, with $G$, defines
the ``MOND radius''
\beq \rM\equiv  \left(\frac{\mb G}{\az}\right)^{1/2}=\left(\frac{r\_s\ell\_M}{2}\right)^{1/2},   \eeqno{mondradius}
where $r\_s$ is the Schwarzschild radius of $\mb$, and $\ell\_M$ the MOND length, which is comparable with the Hubble distance.
It also defines the ``MOND velocity''
\beq \vinf\equiv (\mb G\az)^{1/4}.  \eeqno{vinf}
MOND also defines a universal (i.e., mass independent) fiducial surface density
\beq \SM\equiv\frac{\az}{2\pi G}.  \eeqno{sigmam}
Their special roles in galaxy dynamics have been widely discussed (see below, and the MOND reviews \cite{fm12,milgrom14,milgrom20}).
\par
I note here that $\az$ (with $G$) also defines an important fiducial value of the SAM for a body of mass $\mb$:
\beq \jM(\mb)\equiv\rM\vinf=\mb^{3/4}(G^3/\az)^{1/4},  \eeqno{fidu}
and in terms of $\vinf$
\beq \jM(\mb)=\vinf^3/\az.  \eeqno{fiduvv}
For the canonical value, $\az=1.2~10^{-8}\cmss$ \cite{fm12,milgrom14,milgrom20}, we have
$$\jM(\mb)=383(\mb/10^{10}\msun)^{3/4}\kpc~\kms$$
\beq =269(\vinf/100\kms)^3\kpc~\kms.  \eeqno{numar}
\subsection{Role of $\jM$ in scaling relations \label{scaling}}
To better understand the role of $\jM$ we need to recapitulate the special roles of $\rM$ and $\vinf$.
The MOND radius, $\rM$, plays an important role in defining the transition radius from the Newtonian to the deep-MOND regime. Bodies whose effective size is much larger than their $\rM$ are governed by deep-MOND, scale-invariant dynamics, while bodies contained well within their $\rM$ are, by and large, governed by standard dynamics (Newtonian or general relativistic).
\par
While there are objects (such as stars) that are much smaller than their MOND radius,
galaxies of all types are not found to be appreciably smaller than their MOND radius. This is the expression in MOND of early observations dubbed the ``Freeman law'' (for disc galaxies) and ``Fish law'' (for elliptical galaxies), which state that the distribution of mean surface densities, $\avS$, of galaxies is cut off above a certain value, $\avS_0$ (see, e.g. Ref. \cite{mbs95}, in particular, their Fig. 3).\footnote{The original statements where that $\avS$ was narrowly distributed around some $\avS_0$. We now understand  that this was due to selection against galaxies with $\avS<\avS_0$, of which there are, indeed, many.} With the advent of MOND it was pointed out that this value $\avS_0$ corresponds to a mean acceleration ($G\avS_0$) of $\approx\az$. Moreover, there are explanations for the existence of such a cutoff in the framework of MOND (e.g., Refs. \cite{milgrom83b,milgrom84,milgrom89,brada99a}).
\par
$\vinf$ is of special significance in that MOND predicts that the rotational speed on circular orbits around a mass $\mb$ should become independent of orbital radius at asymptotically large radii, leveling at the value $\vinf(\mb)$.
But, $\vinf$ features not only on the asymptote, it also sets the velocity scale for the whole galaxy. For example, for self-gravitating bodies that are deep in the MOND regime (i.e., of mean accelerations $\ll\az$) MOND predicts that the mass-weighted root-mean-square velocity is $\av{V^2}^{1/2}\approx 0.8\vinf$ (see, e.g., Refs. \cite{milgrom12,milgrom14a} for the predictions, and Refs. \cite{mm13a,pl20} for observational verification).
\par
At the other end, for high-surface-density galaxies, barring some extreme examples, the rotational speed remains roughly constant at $\vinf$ down to small radii. This is understood in MOND as tantamount to the existence of the Freeman cutoff:
Without further knowledge, and based only on AM argumentation, a galaxy can, in principle, be envisaged that has $\avv\gg\vinf(\mb)$. But this would mean that the bulk of the galaxy is very deep in the Newtonian regime. This however, is largely excluded in MOND by the arguments leading to the Freeman and Fish limits, as indeed is observed in real galaxies. Galaxies with even the maximum rotation speed exceeding $1.5\vinf$ are very rare.
\par
More quantitatively,
the relevant mean of the rotational speeds in the disc of surface density $\S(r)$ is
\beq \avv\equiv\frac{\int\S(r)r^2V(r)dr}{\int\S(r)r^2dr}, \eeqno{viva}
where $V(r)$ is weighted in the way it enters the SAM of the disc:
\beq j\_D=\frac{\int\S(r)r^2V(r)dr}{\int\S(r)r~dr}=\avr\avv,  \eeqno{kasal}
where the mean disc radius is
\beq \avr\equiv\frac{\int\S(r)r^2dr}{\int\S(r)r~dr}. \eeqno{factor}
The arguments above state that for discs of all types we have in MOND $\avv\approx\vinf$.
Even more specifically, I calculated $\avv/\vinf$ for MOND rotation curves of exponential disc of scale length $h$ (for which $\avr=2h$), for different values of $\xi=\rM/h$. I find for all deep-MOND discs (low-surface density), which have $\xi\ll 1$, $\avv/\vinf=0.95$. For  $\xi=1,~2,~4$, I find, respectively, $\avv/\vinf=1.03,~1.12,~1.29$ [with the MOND interpolating function $\n(y)=(1-e\^{-y^{1/2}})^{-1}$, see the MOND reviews \cite{fm12,milgrom14,milgrom20} for its meaning; the exact choice of $\n$ is quite immaterial in the present context].
\par
To recapitulate, to within a factor of order unity, $\vinf$ is the effective rotational speed that enters the SAM {\it of the disc component} in disc galaxies.
 This means that the SAM of these discs is
\beq j\_D\approx \frac{\avr}{\rM}\jM(\mb)=\left(\frac{\SM}{\avS}\right)^{1/2}\jM(\mb),  \eeqno{ipota}
where $\mb$ is the total baryonic mass of the galaxy (including all components, such as a bulge), $\avS=\mb/2\pi\avr^2$ is some mean surface density of the galaxy, and $\SM$ is the MOND surface density defined in Eq. (\ref{sigmam}).\footnote{I adopt for concreteness sake a simplified picture of a disc galaxy of total mass $\mb$, with one rotationally supported disc component of mass $\mb\_D$, and a nonrotating bulge of mass $\mb\_B=\mb-\mb\_D$, with $J$ -- the total AM of baryons, wholly in the disc. Thus, $j\equiv J/\mb$, and $j\_D\equiv J/\mb\_D$.}

\par
Relation (\ref{ipota}) establishes $\jM$ as an underlying cornerstone defining correlations between the AM or SAM of galaxies and other global properties. It does so in ways that depend only little on details of the formation and evolution of these galaxies -- in contradistinction with what happens in the dark-matter paradigm as follows:
\par
High-surface-density galaxies, with few exceptions, have (as per the ``Freeman law'') $\avr\approx\rM(\mb)$. Relation (\ref{ipota}) then tells us that their discs should lie on the fiducial relation $j\_D\approx\jM(\mb)$. Thus, their global SAM should satisfy
$j\approx\jM(\mb)(\mb\_D/\mb)$. In other words, to the extent that high-surface galaxy does not have a dominant, nonrotating bulge it should lie on the $\jM(\mb)$ relation in the $j-\mb$ plane. If it has a substantial bulge it should lie below the relation.
\par
As to low-surface-density galaxies, which are deep in the MOND regime; they are generically of subdominant bulge, and relation (\ref{ipota}) tells us that they should lie above the $\jM(\mb)$ relation; the more so the deeper they are in the MOND regime (i.e. the larger their $\avr/\rM$ is).

\subsection{Role of $\jM$ in galaxy evolution \label{evolution}}
Based on the arguments in the previous sections, MOND can explain why some galaxies become low-surface-density virialized disc galaxy, mostly of negligible bulge, while others virialize into disc galaxies with mean surface densities near the Freeman limit, mostly possessing bulges.\footnote{By ``bulge'' here and in what follows I mean any component
of the galaxy -- possibly also a halo -- that takes up mass but little angular momentum, and
is used solely as a component to be reckoned with in the mass and AM balance. I am not referring to any specific type of bulge or to the possibly complicated way in which it forms, as long as it is a result of a secular evolution, not e.g. of mergers.}
\par
Envisage a protogalactic body\footnote{A ``protogalaxy'' here means the state of the would-be-disc-galaxy when it starts to evolve in isolation. It can be a primordial mass that will develop in isolation, or a pair or a group of galaxies that will eventually merge and evolve into one galaxy whose virialized state we want to determine.} of total mass $\mb$ and total AM $J$ (with SAM $j=J/M$). Suppose that it decouples from its surrounding, and evolves into a virialized state in isolation, and without disintegrating, exchanging mass or AM with neither external entities, nor -- as MOND posits -- with a dark-matter halo.
\par
We saw in Sec. \ref{scaling} that MOND dictates that in the virialized state, the disc has $\avv\approx\vinf(\mb)$, where $\mb$ is the total mass of the galaxy.
Consider first the case with
 \beq \z\equiv j/\jM(\mb)\gg1.  \eeqno{liut}
 If the virialized galaxy possesses a (nonrotating) bulge of mass $\mb\_B$, the disc, of mass $\mb\_D=\mb-\mb\_B$ is left with SAM
$j\_D=j\mb/\mb\_D$, which has to equal $\avr\avv\approx\avr\vinf(\mb)$. To accommodate this the disc must adjust its mean radius to be
\beq \frac{\avr}{\rM(\mb)}\approx \z\frac{\mb}{\mb\_D}\gg 1.   \eeqno{buteq}
The minimal value of $\avr$ is $\z\rM(\mb)$, when no bulge is formed. This state of things describes a low-surface-density disc, as concretely defined in the context of MOND: a disc whose mean radius is much larger than the MOND radius of the galaxy, or equivalently, the mean acceleration in which, $\av{a}=\mb G/\avr^2\ll\az$.
\par
Reference \cite{wkf20} shows results of MOND simulations of collapsing gas clouds (finding that they form virialized disc galaxies with exponential discs). Based on only a handful of simulations they observed that: ``...This  suggests  that  low  surface  brightness galaxies result from collapsing gas clouds with high specific  angular  momentum.'' This is in line with the above general result. It would be interesting to see if this result and the ones below are supported by more quantitative and systematic MOND simulations.
\par
If, on the other hand, the protogalaxy starts with $\z\ll 1$, how can this be accommodated by the disc of the virialized galaxy, given that it has to have $\avv\approx\vinf(\mb)$?
In principle, considering only conservation of mass and AM, the galaxy can end up as a pure disc. But it then has to attain a small enough $\avr$ to make $j\_D=\avr\avv\approx \avr\vinf(\mb)$. This would imply $\avr/\rM\approx\z\ll 1$. So the disc would be deep in the Newtonian regime, and would greatly exceed the Freeman limit. MOND does  not robustly predict that the Freeman limit must be obeyed, but it points to it as the more probable outcome.
\par
If there are processes that limit $\avS$ roughly to the Freeman value, the only route the galaxy can take is to put enough of its mass into a nonrotating bulge and attain the minimal $\avr$ compatible with the Freeman limit $\avr\approx\rM(\mb)$. The disc then sits on the fiducial MOND SAM
$j\_D\approx\jM(\mb)$, which, in turn, dictates that its mass is $\mb\_D\approx \z\mb$, with the bulge mass $\mb\_B\approx (1-\z)\mb$.
Such an enforced formation of a bulge is also in line with the arguments, and compatible with the expected dynamics, leading to a Freeman limit in MOND: When the inner surface density of the disc exceeds the Freeman limit dynamics become Newtonian, hence the inner disc becomes unstable, not being protected by the stabilizing MOND effects \cite{milgrom83b,milgrom89,brada99a}, hence is amenable to churning up a bulge, leaving the galaxy at the Freeman limit.

\section{Comparison with observations \label{observation}}
The expectations argued for in Sec. \ref{scaling} call for plotting the disc SAM, $j\_D/\jM(\mb)$, against the mean surface density, $\avS$. MOND leads us to expect that for $\avS$ at the high end -- i.e., near the Freeman limit -- $j\_D$ should be given by the MOND fiducial, $\jM$, calculated for the total mass of the galaxy. But, as we go to progressively lower $\avS$ values, below the MOND value,  $j\_D$ should be above the $\jM(\mb)$ line and progressively depart from it.
\par
I could not find in the literature plots of exactly this desired description, but there are to be found analyses plotting $j$ against other global properties of the galaxies, which can be used as reasonable proxies.
\par
\begin{figure}[ht]
	\centering
\includegraphics[width = 8cm] {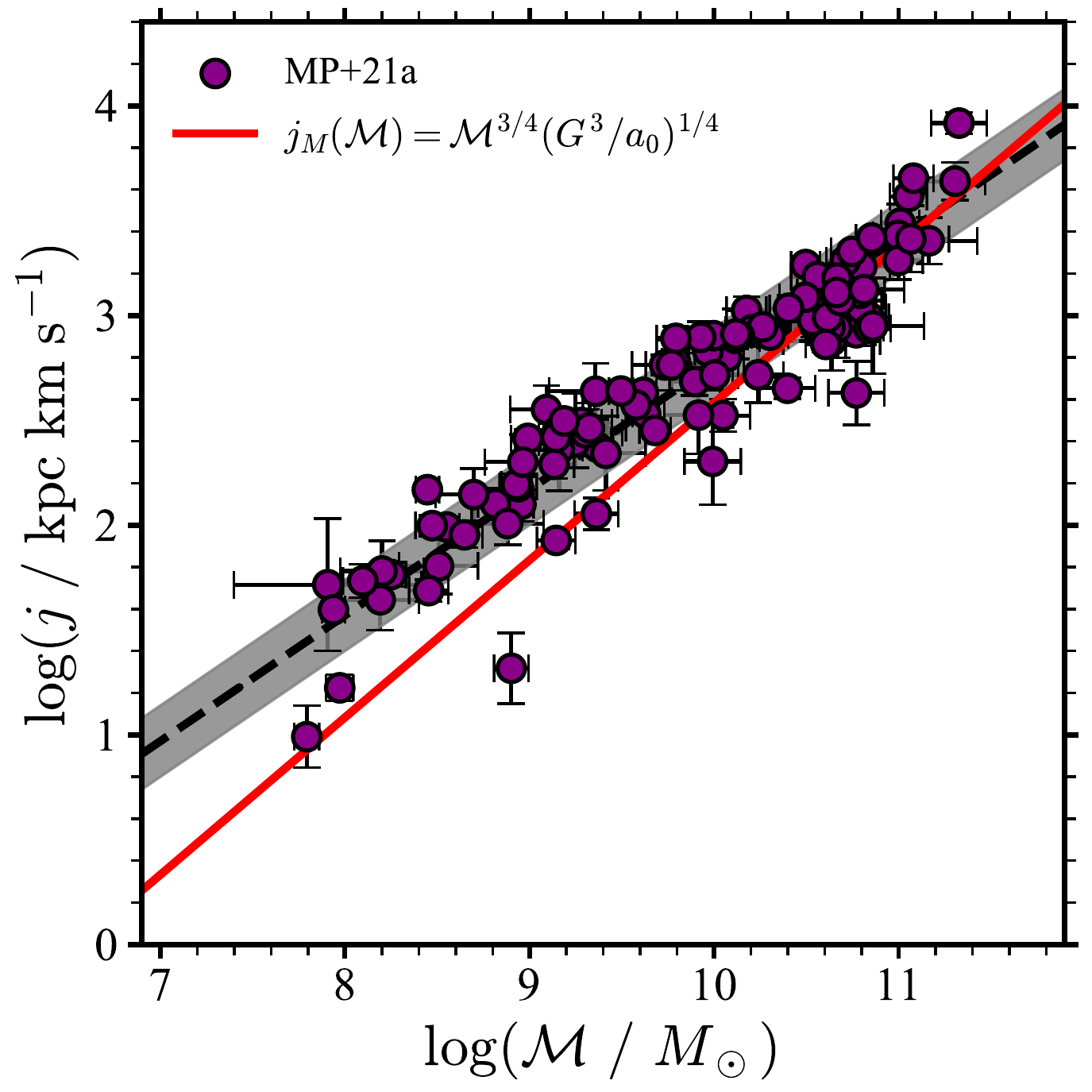}
\caption{The baryonic SAM against the total baryonic mass. Data for a sample of disc galaxies from Ref. \cite{pina21a} are shown, with their best fit power law (dashed line). The thick (red) line shows the fiducial MOND relation $\jM(\mb)$ from Eq. (\ref{numar}). Figure, courtesy of Pavel Mancera Pi\~{n}a.}		\label{asa}
\end{figure}

I chose for comparison the recent analysis of Ref. \cite{pina21a} who plotted, for a sample of disc galaxies the global baryonic SAM ($j=J/\mb$ in our notation) against $\mb$. Their points are reproduced here in Fig. \ref{asa}, together with the MOND $\jM(\mb)$.
\par
The mean surface densities of discs are correlated with their masses (hence with $\vinf\propto \mb^{1/4}$), in that, by and large, low-surface-density galaxies are of lower mass (see, e.g., Fig. 2 of Ref. \cite{lms16}, for the stellar quantities, ignoring the gas) \footnote{There are relatively rare examples of high-mass, low-surface-density galaxies.}. In plots of $j$ vs. $\mb$ we thus expect high mass galaxies to lie around the $\jM(\mb)$ line, rising above it for smaller and smaller $\mb$ values.
\par
Progressively less massive disc galaxies are also becoming progressively low in surface density. With this in mind we see from Eq. (\ref{ipota}) that the galaxies in the sample at the higher-mass end, which have $\avS\approx\SM$,  do indeed fall around the $\jM$ line, while at lower and lower masses, with progressively smaller $\avS$, they do, as expected, fall above the $\jM$ line, and depart progressively from it.
\par
The MOND fiducial $\jM(\mb)$ thus predicts correctly the amplitude of the observed relation as defined by the high-mass end, and the qualitative behavior with galaxy mass.

\section{Discussion \label{discussion}}
I have explained how the fiducial MOND SAM, $\jM$ plays a central role in defining the properties of virialized disc galaxies, and how
it defines the borderline in the mass-AM plane of protogalaxies between those that will become bulgeless low-surface-density discs, and those that will possess a bulge and have mean surface densities near the Freeman limit.
\par
The former of the two roles does not depend much on formation scenarios, as it concerns, and is based on, MOND dynamics of settled galaxies that follow from only the basic tenets of MOND. Its application requires, though, considerations of several caveats. For example, we used simplistic picture of a disc galaxy as made of a single component disc, and a nonrotating bulge. Galactic discs generally have a stellar and a gas component with different masses, and different surface densities, hence different SAM. (They both have, however, roughly the same $\avv\approx\vinf(\mb)$.)
This is not a concern at the high-mass end, where the gas disc is by far subdominant generically. Generically the gas fraction increases with decreasing surface density (and mass) (see, e.g., Fig. 4 of Ref. \cite{lms16}, which, however use the surface brightness, not the total surface densities -- including gas -- which are relevant here) and at the low end we may thus be dealing again with a single dominant gas component, which generically has a different $\S(r)$ than the stars (as in the galaxies studied in Ref. \cite{sanders21}). One should then be careful not to use $\avr$ and $\avS$ values based on the stellar distribution alone.
\par
References \cite{jb19,pina21a}, among others, also plot separately the SAM of gas and stars as functions of their respective masses. But the interpretation of these based on $\jM(\mb)$ would be rather roundabout, for example because it would have to involve the correlation of the gas fraction with total mass, which is much a matter of evolution, and not cleanly an aspect of the dynamics.
\par
I thank Pavel Mancera Pi\~{n}a for preparing the figures, and an anonymous referee for useful comments.

\begin{figure}[ht]
	\centering
\includegraphics[width = 9cm] {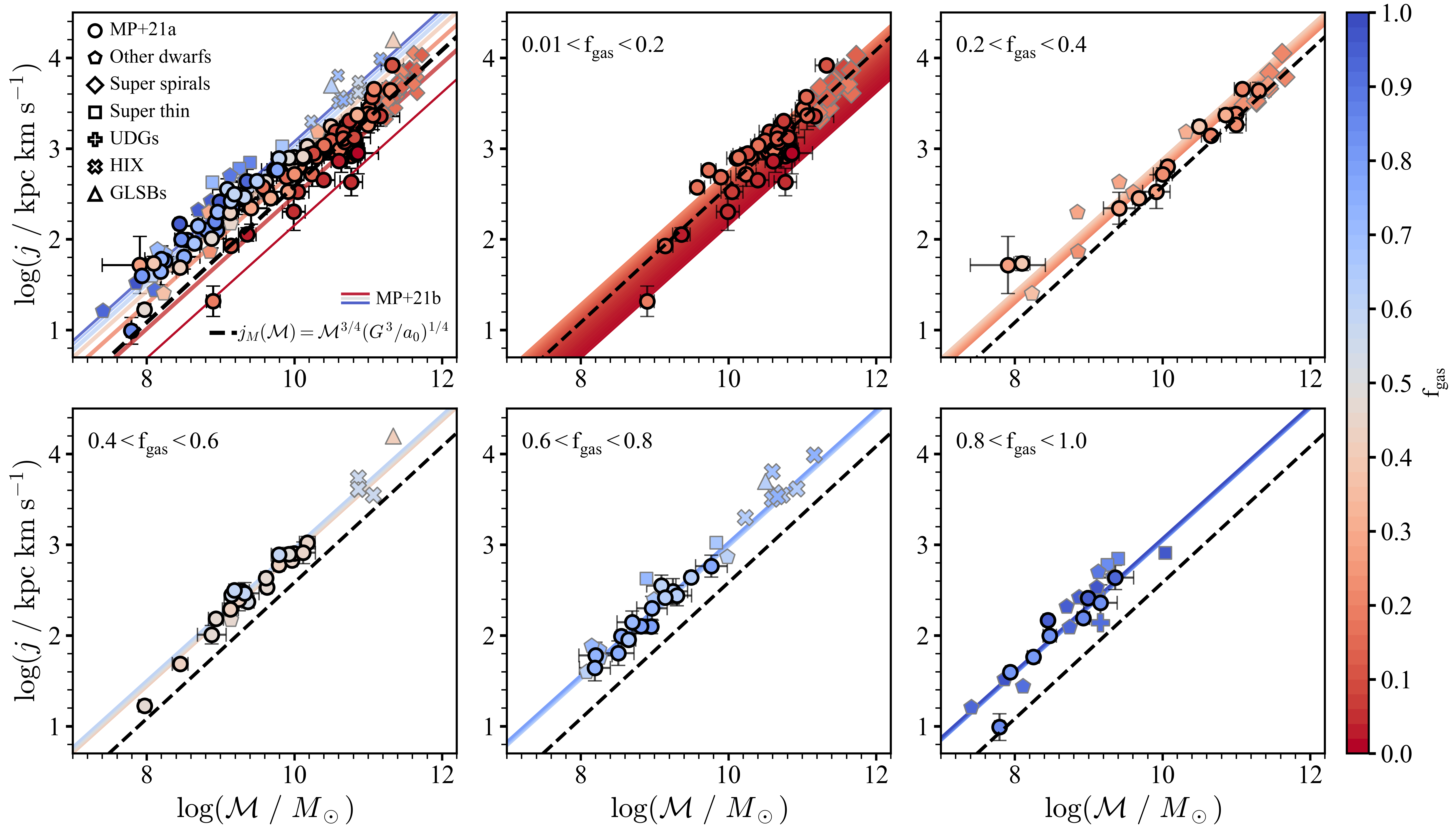}
\caption{The $j-\mb$ relations for a sample of disc galaxies plotted separately for different bins of the gas-mass fraction, shown each with the best-fit power law in solid (colored) lines, taken from Ref. \cite{pina21b}. Also shown in each panel is the fiducial MOND relation $\jM(\mb)$ from Eq. (\ref{numar}) (dashed line). The upper-left panel shows all galaxies on the same $j-\mb$ plot, with all the best-fit lines also shown (and color coded). Color coding for the gas fraction is given on the right. Figure, courtesy of Pavel Mancera Pi\~{n}a.}		\label{baasa}
\end{figure}

{\it Note added.--} After completion of this paper there appeared an analysis of disc-galaxy $j-\mb$ relations, with results that lend themselves much better to direct comparison with the MOND predictions. Reference \cite{pina21b}, using an even larger sample of disc galaxies than in Ref. \cite{pina21a}, plot the $j-\mb$ relations separately for galaxy subsamples characterized by different values of the gas fraction. These plots are reproduced in Fig. \ref{baasa}, together with the fiducial, MOND $\jM(\mb)$ relation. The main points to note are
\begin{enumerate}[(i)]
\item
For each gas-fraction bin, the observed $j-\mb$ relation is parallel to the MOND fiducial relation (with a slope of $\approx 3/4$). Inasmuch as the gas fraction is a proxy of the mean surface density, this agrees with the MOND prediction for a constant surface density [Eq. (\ref{ipota})].
\item
The higher the gas fraction -- corresponding to progressively lower mean surface density -- the higher is the observed plot above the fiducial relation -- also as predicted in Eq. (\ref{ipota}). A quantitative measure of the actual mean surface densities for each bin would be desirable for quantitative comparison with the MOND prediction, but the trend is as predicted by Eq. (\ref{ipota}).
\item
In the highest-gas-fraction bin (lower-right panel) the data points lie above the fiducial relation by about half an order of magnitude. According to Eq. (\ref{ipota}), this would correspond to a surface density $\sim 10$ times lower than the Freeman limit.\footnote{Note that what is relevant here are mass surface densities. So, for the high-gas-fraction galaxies it is by and large the gas surface density that is relevant, while for the low-gas-fraction galaxies it is by and large the stellar surface density.}
\item
The galaxies in the lowest gas fraction bin (upper-middle panel), which are at the high-surface-density end, fall on the fiducial MOND relation, showing that their mean surface densities are at the Freeman value, which corresponds to the MOND $\SM$, defined in Eq. (\ref{sigmam}).
\item
The upper-left panel shows all galaxies on the same plot. This is in the same vein as Fig. \ref{asa}. It shows clearly why the slope of the full-sample correlation is shallower than that of the fiducial MOND relation: As explained above, in Sec. \ref{observation}, this results from galaxies with lower masses having typically higher gas fractions (lower surface density) hence a higher normalization of the $j-\mb$ relation relative to the fiducial.
\end{enumerate}
It emerges then that MOND all but removes the mystery from the $j-\mb$ relation, whose origin no longer has to be looked for in complicated formation scenarios, as in the dark-matter paradigm.

\end{document}